\documentclass[aps,prb,twocolumn,superscriptaddress,showpacs,floatfix]{revtex4}

\usepackage{graphicx,graphics,color,float}
\usepackage{dcolumn}
\usepackage{bm}
\usepackage{stmaryrd}
\usepackage{latexsym}
\usepackage{amssymb}
\usepackage{amsfonts}
\usepackage{amsmath}

\begin{document}

\title{Pressure induced topological quantum phase transition in Sb$_2$Se$_3$}

\author{Wei Li}
\affiliation{Shanghai Center for Superconductivity and State Key Laboratory of Functional Materials for Informatics, Shanghai Institute of Microsystem and Information Technology, Chinese Academy of Sciences, Shanghai 200050, China}
\affiliation{Department of Physics, State Key Laboratory of Surface Physics and Laboratory of Advanced Materials, Fudan
University, Shanghai 200433, China}

\author{Xin-Yuan Wei}
\affiliation{Department of Physics, State Key Laboratory of Surface Physics and Laboratory of Advanced Materials, Fudan
University, Shanghai 200433, China}

\author{Jian-Xin Zhu}
\affiliation{Theoretical Division and Center for Integrated Nanotechnologies, Los Alamos National Laboratory, Los Alamos, New Mexico 87545, USA}

\author{C. S. Ting}
\affiliation{Texas Center for Superconductivity and Department of Physics, University of Houston, Houston, Texas 77204, USA}

\author{Yan Chen}
\affiliation{Department of Physics, State Key Laboratory of Surface Physics and Laboratory of Advanced Materials, Fudan
University, Shanghai 200433, China}

\date{\today}


\begin{abstract}
Based on the first-principles band structure calculations, we investigate the effects of hydrostatic pressure on the conventional insulator (CI) Sb$_2$Se$_3$ and predict that it undergoes a topological quantum phase transition from a CI to a non-trivial topological insulator at a critical pressure value. The pressure induced topological quantum phase transition is confirmed by calculating the evolution of the bulk energy gap as a function of pressure, the inversion of energy band structure and the $Z_2$ topological invariant as well as the existence of the Dirac-like topological surface states. Our predictions can be tested by both spectroscopy and transport experiments.
\end{abstract}

\maketitle

\section{Introduction}
\label{Intro}

Over the years, there have been significant theoretical and experimental efforts from condensed matter and materials physics communities to understand and identify the topological quantum phase transition (TQPT) as well as different topological phases of matter~\cite{XGWen,Wilczek}. These topological phases can be characterized by various kinds of topological properties such as topological invariants, non-trivial topological protected edge/surface states and/or unusual elementary excitations
in the bulk or surface. A more recent promising realization of topological phases of matter is the time reversal invariant (TRI) topological insulators (TIs)~\cite{MZHasan,XLQi}.

The TIs in two~\cite{BHZ,SCZhang2007} or three~\cite{LFuPRL2007,JEMoore,RRoy,DHsieh2008,HZhangNP,YXia2009,DHsieh2009,YLChen,DHsiehPRL,WZhang,CChen} dimensions have a bulk energy band gap, like a conventional insulator (CI), but have a conducting gapless edge or surface states protected by time reversal symmetry. According to Kramer's theorem, the edge or surface states consisting of an odd number of massless Dirac cones are robust against TRI perturbations, while those states consisting of an even number of massless Dirac cones are not protected. This even-odd effect is the key reason why TIs are characterized by a $Z_2$ topological invariant~\cite{CLKane,LFuPRL2007}. Interestingly, a uniaxial pressure in Sb$_2$Se$_3$ induced TQPT has been proposed theoretically~\cite{WLLiu}. But it is very hard to realize experimentally that is because of the layer structure of Sb$_2$Se$_3$ resulting a van der Waals-type coupling between each quintuple layers (QLs). Thus, both the theoretical and experimental studies of control parameters induced TQPT are still awaiting.

Recently, Bera \textit{et al.}~\cite{ABera} observed a hydrostatic pressure induced transition in Sb$_2$Se$_3$ using Raman spectroscopy and obtained a bulk energy band gap closing at the critical pressure value using the first-principles calculations. That work indicated this system undergoes a TQPT from a CI phase to a non-trivial TI one with reversal of parity of electronic bands passing through a metallic state at the transition point by tuning the pressure. However, the mechanism of the hydrostatic pressure induced TQPT observed by Raman spectroscopy~\cite{ABera} remains unclear. From the viewpoint of the band structure calculations in a well-understood framework, to undergo such a TQPT, four criteria must be satisfied: (i) The bulk energy gap must be closed when the TQPT occurs; that is, when the pressure reaches the critical value P$_c$, the bulk energy gap should vanish. (ii) The conduction and valence bands cross, forming a \textquotedblleft band inversion\textquotedblright\ at the TRI $\mathbf{k}$ point in the Brillouin Zone (BZ) at a pressure above P$_c$. (iii) The $Z_2$ topological invariant should have a different number in the two phases divided by the bulk energy gap closing point. (iv) The gapless Dirac-like topological surface states inside the bulk energy gap only exist in one phase, while the other one does not exist. It is important to note that the criterion (iii), that the $Z_2$ topological invariant changed, suffices to understand the nature of occurring a TQPT and it also implies the others criteria based on the context of band theory and the idea of the bulk-surface correspondence of the TIs~\cite{Hatsugai}.

In this paper, we theoretically study the hydrostatic pressure induced TQPT in Sb$_2$Se$_3$ motivated by aforementioned discussions of four fundamental criteria, as well as the mechanism for TQPT based on the first-principles calculations. Our results show that when the pressure is increased to the critical value, the bulk energy gap is closed accompanied by the TQPT occurring. Our results also predict that a \textquotedblleft band inversion\textquotedblright\ appears at the TRI $\Gamma$ point in the BZ at a pressure value above P$_c$, which is attributed to the pressure induced enhancement of the crystal field splitting, resulting in the crossing between conduction and valence bands. This mechanism of realizing TQPT is qualitatively the same as previous theoretical proposal by using the uniaxial pressure~\cite{WLLiu}. To quantitatively confirm the TQPT occurring at the critical pressure value, we calculate the $Z_2$ topological invariant in two phases separated by the bulk energy gap closing point and show that the low pressure phase has a trivial $Z_2$ topological invariant number $(0;000)$, while the high pressure phase has a non-trivial $Z_2$ topological invariant number $(1;000)$. According to the idea of the bulk-edge correspondence of the TIs, we also calculate the surface spectral function and show that the Dirac-like topological surface states inside the bulk energy gap only exist in the high pressure phase, a non-trivial strong TI phase. The experimental validation of our theoretical predictions should be readily achievable by both spectroscopy and transport measurements.

The structure of this paper is as follows. In Sec.~\ref{discussion}, we present the theoretical results and discussions. Finally, in Sec.~\ref{conclusion}, we summarize our main conclusions.

\begin{figure}[t]
\includegraphics[width=8.8cm, height=4.5cm]{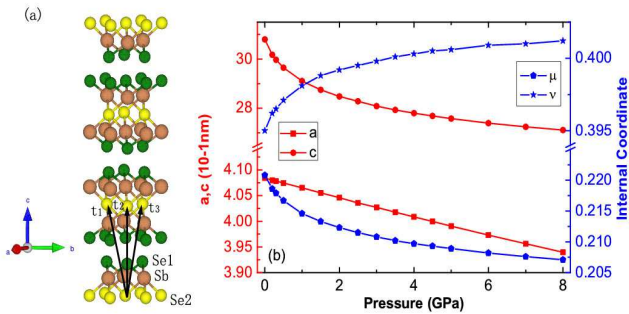}
\caption{(Color online) (a) Crystal structure of the Sb$_2$Se$_3$ and the primitive translation vectors $\mathbf{t}_{1,2,3}$ are shown. (b) The pressure dependent optimized lattice constants and the internal coordinates of the Sb$_2$Se$_3$, where the Se2 site is set to be at the origin $(0, 0, 0)$, the two Sb  sites are at $(\pm\mu, \pm\mu, \pm\mu)$, and two Se1 sites are at $(\pm\nu, \pm\nu, \pm\nu)$, defined in the unit of primitive translation vectors, as shown in (a).}
\label{fig:fig1}
\end{figure}

\section{Results and discussions}
\label{discussion}

The first-principles calculations implemented in the VASP code~\cite{VASP}, the plane wave basis method, and the ultrasoft pseudopotential method with generalized gradient approximation (USPP-GGA) exchange correlation potential have been used. Additionally, the spin-orbit coupling is also included through all over the calculations. A 500eV cutoff in the plane wave expansion and a $12\times 12\times 12$ Monkhorst-Pack $\mathbf{k}$-grid are chosen to ensure the calculation with an accuracy of $10^{-5}$eV, and all structures (lattice constants as well as internal coordinates) were optimized until forces on individual atoms were smaller than 0.005eV/\AA\ to obtain sufficient accuracy throughout the calculations.

\begin{figure}[t]
\includegraphics[width=9.5cm, height=6.5cm]{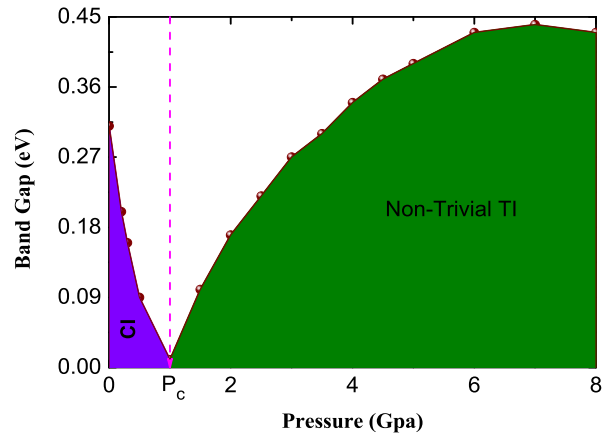}
\caption{(Color online) The phase diagram of the bulk energy band gap as a function of the pressure. The magenta-dashed line denotes the critical point pressure P$_c\simeq 1.0$ GPa, when the bulk energy gap is closed. The low pressure phase is a CI, while the high pressure phase is a non-trivial TI.}
\label{fig:fig2}
\end{figure}

Sb$_2$Se$_3$ is assumed to have a rhombohedral crystal structure~\cite{ABera} with space group $D_{3d}^5(R\bar{3}m)$, similar to that of Bi$_2$Se$_3$ and shown in Fig.~\ref{fig:fig1}(a). This system contains QLs ordered in a Se1-Sb-Se2-Sb-Se1 sequence along the $\hat{\mathbf{c}}$-axis, and the primitive translation vectors $\mathbf{t}_{1,2,3}$ are: $\mathbf{t}_1=(-a/2,-\sqrt{3}a/6,c/3)$, $\mathbf{t}_2=(a/2,-\sqrt{3}a/6,c/3)$, and $\mathbf{t}_3=(0,\sqrt{3}a/3,c/3)$, respectively, where $a$ and $c$ are lattice constants in the hexagonal cell. The internal coordinates of atoms Se1, Se2, and Sb are set to be at the $(\pm\nu,\pm\nu,\pm\nu)$, $(0,0,0)$, and $(\pm\mu,\pm\mu,\pm\mu)$, respectively. We optimized both the lattice constants and internal coordinates under the hydrostatic pressure ranged from $0$ GPa to $8$ GPa without considering the crystal structure phase transition~\cite{JLZhang,CZhang,ABera} occurring. If the pressure increases to a much higher value, the crystal structure will change and the TIs could become conductors or CIs. The issue of the structure stability of Sb$_2$Se$_3$  will be left to be addressed in future studies. Those optimized results are all shown in Fig.~\ref{fig:fig1}(b). At ambient pressure P$=0$ GPa the optimized lattice constants for Sb$_2$Se$_3$ are consistent with previous results~\cite{HZhangNP}. From Fig.~\ref{fig:fig1}(b), we notice that the lattice constants $a$ and $c$ are shrunk with increasing pressure, as one expects intuitively. However, the internal coordinate of the Sb atom increases with higher pressures, which seems to be counterintuitive. This feature can be understood from the binding energy consideration: the coupling is strong between two atomic layers within one QL, while the coupling between QLs, which is predominately a van der Waals-type interaction, is much weaker. Consequently, the lattice coordinate of the Sb atom increases to minimize the total energy.

Next, we turn to discuss the relationship between pressure and the bulk energy gap, as shown in Fig.~\ref{fig:fig2}. The calculated bulk energy gap for Sb$_2$Se$_3$ at ambient pressure is $0.31$ eV, which is in good agreement with previous experimental and computational results (about $0.2 \sim 0.3$ eV)~\cite{HZhangNP,Black,Mooser}. With increasing pressure, the bulk energy gap is gradually reduced. Eventually, the bulk energy gap is closed when the pressure reaches to the critical value P$_c \simeq 1.0$ GPa. It is worth to point out that the calculated critical value P$_c\simeq 1.0$ GPa using USPP-GGA is a bit smaller than the experimental value of P$_c^e$$ \approx$ 1.6$\sim$2.5 GPa~\cite{ABera}, which difference originates mainly from the fact that lattice constants and bond lengthes optimized by the pseudopotential method deviate slightly  from the experimental values. We also performed extensive calculations with various pseudopotentials and the main features of TQPT remain qualitatively unaltered. Due to the fact that the spin-orbit coupling strength in Sb$_2$Se$_3$ is not strong enough to induce the band inversion at ambient pressure, Sb$_2$Se$_3$ is maintained as a CI phase at pressures up to the P$_c$. If the pressure is increased beyond the P$_c$, the bulk energy gap will be reopened and the system returns to an insulating phase. The bulk energy gap can reach up to a maximal value $E_g \simeq 0.44$ eV with increasing pressure, which is much larger than the energy scale of room temperature. Such a gap closing-reopening transition by tuning the pressure indicates that there is a phase transition occurring at the critical point P$_c$.

\begin{figure}[t]
\includegraphics[width=8.5cm, height=6.5cm]{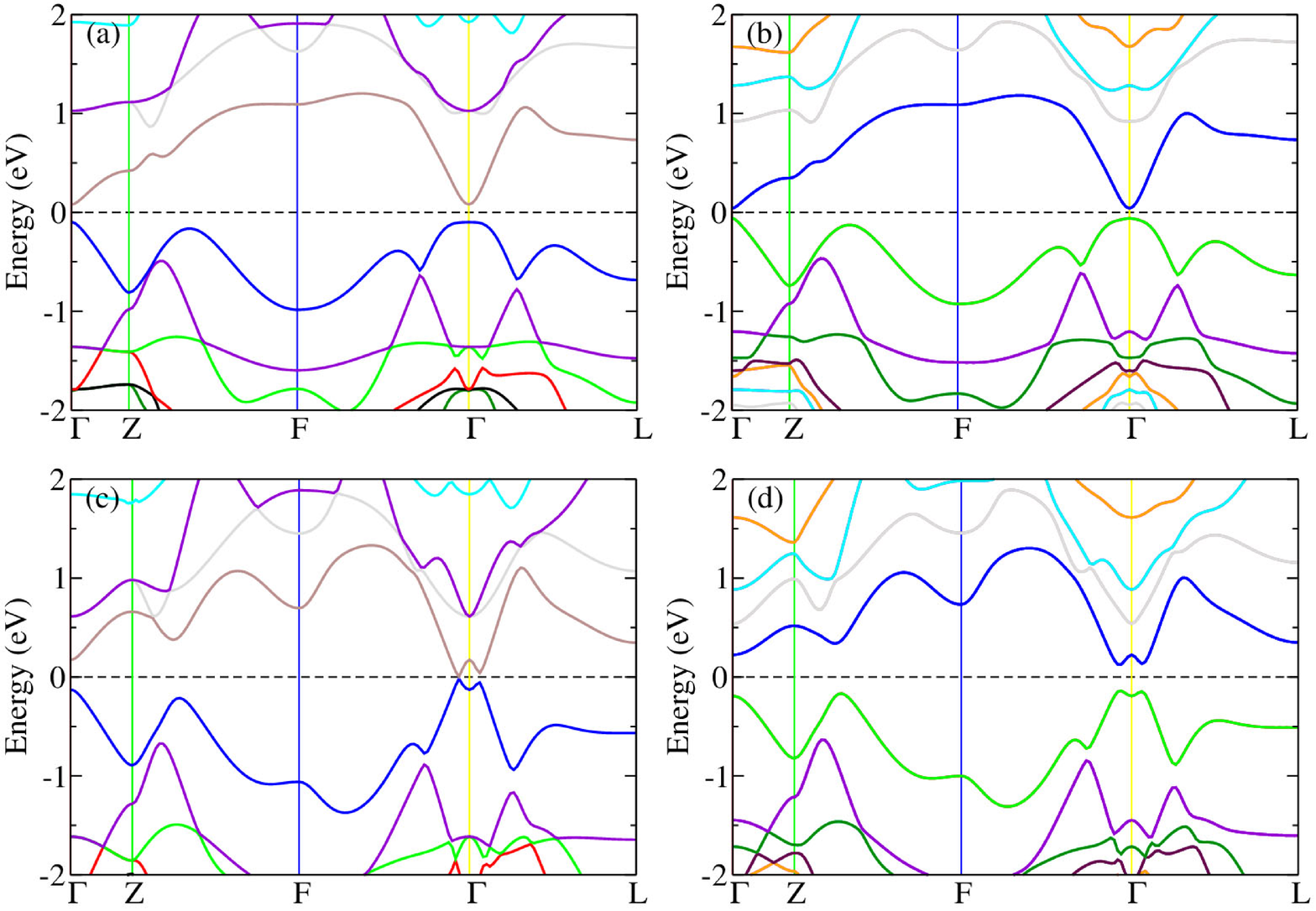}
\caption{(Color online) Energy band structure for Sb$_2$Se$_3$ under pressures of P$=0$ GPa (a)-(b) and P$=3.0$ GPa (c)-(d). The spin-orbit coupling is taken into account in (b) and (d). The dashed line indicates the Fermi level.}\label{fig:fig3}
\end{figure}

To understand the nature of the new phase, we calculate the bulk energy band structure. The energy band structure with and without spin-orbit coupling at ambient pressure and the pressure value with P$=3.0$ GPa are shown in Fig.~\ref{fig:fig3}. Comparing with Fig.~\ref{fig:fig3}(a) and (b), we demonstrate that the inclusion of spin-orbit coupling does not lead to a band inversion at ambient pressure. Further comparing the ambient pressure data with the P$=3.0$ GPa results shown in Fig.~\ref{fig:fig3}(b) and (d), we notice that there is an crossing feature around the $\Gamma$ point in Fig.~\ref{fig:fig3}(d) with the P$=3.0$ GPa, which indicates an inversion between the conduction band and valence band, suggesting that the new phase in Sb$_2$Se$_3$ under high pressure is a non-trivial TI phase. Therefore, we demonstrate that the pressure induced phase transition is indeed a TQPT from a CI phase to a non-trivial TI one.

\begin{table}
\caption{Parities $\delta_i$ at eight TRI $\mathbf{k}$ points for Sb$_2$Se$_3$ under different pressure values. The $Z_2$ topological invariants are $(1;000)$ for a pressure value above P$_c$ while the $Z_2$ topological invariants are $(0;000)$ for a pressure value below P$_c$, which indicate a non-trivial strong TI and a CI, respectively.}\label{tableParity}
\begin{ruledtabular}%
\begin{tabular}{cccccccccc}
 & $\delta_{1}$ & $\delta_{2}$  & $\delta_{3}$  & $\delta_{4}$  & $\delta_{5}$  & $\delta_{6}$  & $\delta_{7}$  & $\delta_{8}$  & ($\nu_{0};\nu_{1}\nu_{2}\nu_{3})$\tabularnewline
\hline
0GPa  & +1  & +1  & +1  & +1  & +1  & +1  & +1  & +1  & (0;000)\tabularnewline
0.8GPa& +1  & +1  & +1  & +1  & +1  & +1  & +1  & +1  & (0;000)\tabularnewline
1.2GPa  & -1  & +1  & +1  & +1  & +1  & +1  & +1  & +1  & (1;000)\tabularnewline
3GPa  & -1  & +1  & +1  & +1  & +1  & +1  & +1  & +1  & (1;000)\tabularnewline
\end{tabular}\end{ruledtabular}
\end{table}

We now proceed to elaborate the mechanism of the pressure induced TQPT. It is known from the atomic physics that the spin-orbit coupling strength is proportional to $Z^4$ (where $Z$ is the atomic number; $Z=51$ for Sb and $34$ for Se). As we have shown above, the spin-orbit coupling strength in Sb$_2$Se$_3$ is not strong enough to induce the band inversion at ambient pressure. Therefore, the spin-orbit coupling is unable to induce the band inversion even under a high pressure value and it cannot explain the Raman experimental results~\cite{ABera} either. To explore the mechanism of TQPT, we start from the crystal field splitting without considering the spin-orbit coupling effect. According to the point-group symmetry, the $p_z$ orbital is split from the $p_x$ and $p_y$ orbitals and the latter of the two orbitals remains degenerate on the Sb and Se atoms. Due to such splitting, the energy levels around the Fermi energy forming the valence and the conduction bands exhibit a large energy gap. This gap is much larger than the energy scale of the spin-orbit coupling at ambient pressure (comparing with Fig.~\ref{fig:fig3}(a) and (b)). When the pressure is applied, the crystal structure will be squeezed. As a result, the level splitting is enhanced and the bulk energy gap between the top of valence band and the bottom of conduction band at $\Gamma$ point is gradually reduced. If the pressure is increased enough, the valence band and the conduction band will cross to form a semi-metallic state, see Fig.~\ref{fig:fig3}(c). In Fig.~\ref{fig:fig3}(c), there are two accidental degenerate states appearing around the $\Gamma$ point, which are not protected by time reversal symmetry. Thus, if we take the spin-orbit interaction into account, these accidental degenerate states will be broken and a non-trivial TI is realized (see Fig.~\ref{fig:fig3}(d)).

\begin{figure}[tbp]
\includegraphics[width=8.5cm, height=7.5cm]{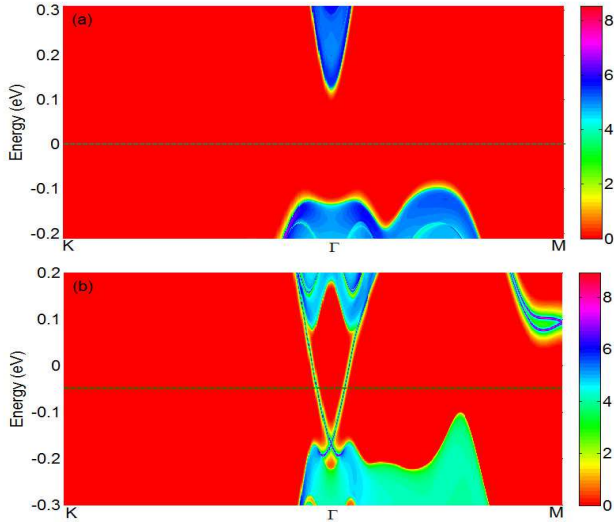}
\caption{(Color online) Energy and momentum dependence of the surface spectral function  for Sb$_2$Se$_3$ at ambient pressure (a) and the pressure value P$=3.0$ GPa (b) on the $[111]$ surface. The red regions indicate bulk energy gaps while the green regions indicate bulk energy bands. The Dirac-like topological surface states can only be seen clearly around the $\Gamma$ point under the pressure P$=3.0$ GPa (lower panel). The dashed line indicates the Fermi level.}
\label{fig:fig4}
\end{figure}

To firmly confirm that the TQPT is induced by pressure, we further calculate the $Z_2$ topological invariant. Because of the existence of the spatial inversion symmetry in Sb$_2$Se$_3$, the calculation of the $Z_2$ topological invariant can be dramatically simplified by the so-called \textquotedblleft parity method\textquotedblright~\cite{LFuPRB2007}. Accordingly, the $Z_2$ topological invariant of Sb$_2$Se$_3$ under pressure can be obtained from the wave function parities at the eight TRI $\mathbf{k}$ points defined as $\Gamma_{i=(n_1n_2n_3)} = \frac{1}{2}(n_1\mathbf{G}_1 + n_2\mathbf{G}_2 + n_3\mathbf{G}_3)$, where $\mathbf{G}_j$ are primitive reciprocal lattice vectors with $n_j = 0$ or $1$, and determined by the quantities
$\delta_i = \prod_{m=1}^{N_{occ}}\xi_{2m}(\Gamma_i)$.
Here, $\xi_{2m}(\Gamma_i) = \langle \Phi_{2m,\Gamma_i}|\hat{P}|\Phi_{2m,\Gamma_i}\rangle=\pm 1$ is the eigenvalue of parity operator $\hat{P}$, which corresponds to even (odd) parity of the Bloch functions $|\Phi_{2m,\Gamma_i}\rangle$ at the $2m$-th occupied band and the TRI $\mathbf{k}$ point $\Gamma_i$ $(i=1,..,8)$. The product is taken over all of the occupied bands $N_{occ}$ with only even band index due to the Kramers degeneracy $\xi_{2m} = \xi_{2m-1}$ at the TRI $\mathbf{k}$ point. In Sb$_2$Se$_3$ system, there are four independent invariants $(\nu_0;\nu_1\nu_2\nu_3)$, $(-1)^{\nu_0} = \prod_{i=1}^{8}\delta_i$, and $(-1)^{\nu_k} = \prod_{n_k=1,n_{j\neq k}=0,1}\delta_{i=(n_1n_2n_3)}$,
where $\nu_0$ is the strong topological invariant and independent of the choice of primitive reciprocal lattice vectors $\mathbf{G}_j$, while $\nu_1$, $\nu_2$, and $\nu_3$ defined as weak topological invariants are not. The combination of these four independent invariants ($\nu_0;\nu_1\nu_2\nu_3$) clearly distinguish three classes of states: non-trivial TI with either strong TI or weak TI, and CI. The calculated parities $\delta_i$ $(i=1,...,8)$ at the eight TRI $\mathbf{k}$ points with relative coordinates $(0, 0, 0)$, $(0, 0, 0.5)$, $(0, 0.5, 0)$, $(0, 0.5, 0.5)$, $(0.5, 0, 0)$, $(0.5, 0, 0.5)$, $(0.5, 0.5, 0)$, $(0.5, 0.5, 0.5)$, respectively, in a primitive reciprocal lattice are listed in Table~\ref{tableParity}. The $Z_2$ topological invariant $(\nu_0;\nu_1\nu_2\nu_3)$ can be obtained by multiplying over the parities $\delta_i$ of all TRI $\mathbf{k}$ points and are also listed in Table~\ref{tableParity}. One can see that the $Z_2$ topological invariants are $(1;000)$ for Sb$_2$Se$_3$ at a pressure value above P$_c$, indicating a non-trivial strong TI, while $Z_2$ topological invariants are $(0;000)$ for Sb$_2$Se$_3$ at a pressure value below P$_c$, demonstrating a CI. This agrees with the above analyzes in that a \textquotedblleft band inversion\textquotedblright\ between the conduction band and valence band occurs at the $\Gamma$ point. It is also shown that a TQPT occurs at the critical pressure value P$_c \simeq 1.0$ GPa.

Finally, we discuss the topological surface states based on the idea of the bulk-edge correspondence of the TIs~\cite{Hatsugai}. In order to reveal the topological features of the Sb$_2$Se$_3$ system under pressure, we calculate the topological surface states of a semi-infinite system with and without the pressure based on the surface Green's function~\cite{Sancho1985,WeiLi} constructed by the maximally localized Wannier function fitting to first-principles band structure calculations using the method developed by Marzari and co-workers~\cite{Marzari1997,Souza2001}. The dispersion of the surface states, as shown in Fig.~\ref{fig:fig4}, was obtained by evaluating the imaginary part of the surface Green's function. Comparing with Fig.~\ref{fig:fig4}(a) and (b), one can see the topological surface states, which form a single Dirac cone at $\Gamma$ point, only exist at the pressure P$=3.0$ GPa. This concludes a non-trivial TI nature and agrees well with the bulk parity analysis and band structures calculations.

\section{conclusion}
\label{conclusion}

In conclusion, we have performed the first-principles calculations for the effect of hydrostatic pressure on the Sb$_2$Se$_3$ and predicted that there is a TQPT occurring at the critical value P$_c \simeq 1.0$ GPa. To verify the TQPT, we firstly calculate the evolution of bulk energy gap as a function of pressure and observe the bulk energy gap closing-reopening transition. Secondly, the band structure calculations show that there is a \textquotedblleft band inversion\textquotedblright\ appearing at the $\Gamma$ point in the BZ at a pressure value above P$_c$, which is attributed to the pressure induced enhancement of crystal field splitting, resulting in the crossing of the conduction and valence bands. Thirdly, the calculated $Z_2$ topological invariant indicates that a CI phase $(0;000)$ and non-trivial strong TI phase $(1;000)$ are separated by the critical pressure value P$_c$. Finally, the calculated Dirac-like topological surface states only exist in the non-trivial strong TI phase $(1;000)$ region. These results can be compared directly with experimental spectroscopy techniques, such as angle-resolved photoemission spectroscopy and scanning tunneling microscopy, and transport measurements. Its verification by future experiments will be a firm step towards understanding and realizing TQPT in real materials.

\section*{Acknowledgement}

We thank R. B. Tao, Z. Fang, X. M. Xie, Z. Liu, J.-H. She, B. Gao, D. W. Shen, G. Xu, and Q. S. Wu  for helpful discussions. This work was supported by the State Key Programs of China (Grant Nos. 2012CB921604 and 2009CB929204) and the National Natural Science Foundation of China (Grant Nos. 11074043 and 11274069) (W.L.\& Y.C.), the National Natural Science Foundation of China No. 61171011 (X.-Y. W.), the Robert A.Welch Foundation under Grant No. E-1146 (C.S.T.), and the National Nuclear Security Administration of the US Department of Energy at LANL under Contract No. DE-AC52-06NA25396 and the US DOE Office of Basic Energy Sciences (J.-X.Z.). W.L. also gratefully acknowledges financial support by the Strategic Priority Research Program (B) of the Chinese Academy of Sciences (Grant No. XDB04010600) and the National Natural Science Foundation of China (Grant No. 11227902).

\end{document}